\documentclass[prb,preprint]{revtex4}
\usepackage{graphicx}
\usepackage{dcolumn}
\usepackage{bm}
\usepackage{amssymb,amsmath}

\begin{document}

\title{Local structural excitations in model glasses}
\author{S. Swayamjyoti}
\author{P. M. Derlet}
\email{Peter.Derlet@psi.ch} 
\affiliation{Condensed Matter Theory Group, Paul Scherrer Institut, CH-5232 Villigen PSI, Switzerland}
\author{J. F. L\"{o}ffler}
\affiliation{Laboratory of Metal Physics and Technology, Department of Materials, ETH Zurich, 8093 Zurich, Switzerland}
\date{\today}

\begin{abstract}
Structural excitations of model Lennard-Jones glass systems are investigated using the Activation-Relaxation-Technique (ART), which explores the potential energy landscape of a local minimum energy configuration by converging to a nearby saddle-point configuration. Performing ART results in a distribution of barrier energies that is single-peaked for well relaxed samples. The present work characterises such atomic scale excitations in terms of their local structure and environment. It is found that, at zero applied stress, many of the identified events consist of chain-like excitations that can either be extended or ring-like in their geometry. The location and activation energy of these saddle-point structures are found to correlate with the type of atom involved, and with spatial regions that have low shear moduli and are close to the excess free volume within the configuration. Such correlations are however weak and more generally the identified local structural excitations are seen to exist throughout the model glass sample. The work concludes with a discussion within the framework of $\alpha$ and $\beta$ relaxation processes that are known to occur in the under-cooled liquid regime.
\end{abstract}
\maketitle

\section{Introduction}

Local structural excitations (LSEs) occurring at the atomic scale of Bulk Metallic Glasses (BMGs) are believed to mediate their unique plastic deformation properties. In the early seminal work of Spaepen~\cite{Spaepen1977}, a free volume model was proposed in which single atoms would migrate to a neighbouring free volume region. Subsequently Argon~\cite{Argon1979} introduced the idea that small groups of atoms undergo a structural transformation that could modify the local shear-stress state. The early atomistic simulations of Falk and Langer~\cite{Falk1998} confirmed this picture and referred to these LSEs as Shear Transformation Zones (STZs), leading to a series of static and dynamical atomistic simulations to better understand how such STZs lead collectively to macroscopic plasticity~\cite{Schuh2003,Maloney2004,Demkowicz2005,Shi2006,Guan2010}. The structural transformations studied in these works were obtained via application of strain rates that are many orders of magnitude higher than that normally seen in experiment, leading to localized material instabilities often identified as STZs. In terms of the traditional thermal activation picture for plasticity, such simulations therefore study the athermal limit of plasticity in bulk metallic glasses.

A common question has been as to whether there exist local structural features within the glass configuration that facilitate the existence of LSEs. Indeed from the early athermal atomistic simulation work, those spatial regions which exhibited a localised material instability and the subsequent observation of STZs have been referred to as pre-existing liquid like regions - spatial domains that are not fully relaxed~\cite{Demkowicz2004,Demkowicz2005a,Demkowicz2005b,Langer2008}. In the work of Langer, a relatively low density of such regions was theoretically considered~\cite{Langer2008}. From this perspective it is then valid to ask whether there are structural features of the unstrained glass configuration that allow identification of regions predisposed to LSEs - the so-called liquid like regions? In the work of L\'{e}onforte {\em et al}~\cite{Leonforte2005}, the reversible non-affine atomic displacements associated with a finite elastic distortion was considered. Such displacements give insight into the spatial extent of the low curvature of the unstrained PEL. This work found a percolative network of large non-affine displacements, however when the applied strain was increased to promote plastic activity, the location of the observed LSEs did not show a strong correlation with the non-affine displacements suggesting the corresponding athermal energy barriers were not directly related to the unstrained PEL curvature. In the work of Mayr \cite{Mayr2009}, who analyzed the local elastic constants (Born plus the kinetic fluctuation contributions) as a function of applied stress and temperature did find a strong connection between emerging elastic instabilities and eventual plasticity as the temperature approached the glass transition. Whilst these simulations were fully dynamical, the necessarily high strain rates place these results close to the athermal regime.

Atomistic simulation methods that avoid the high stress athermal limit, are the so-called potential energy landscape (PEL) exploration methods. The relevance of these methods to bulk metallic glasses relies on the idea that the atomic configuration of a structural glass spends much of its time in a local potential energy minimum, only occasionally transiting it via a saddle-point region to a new local minimum. Such structural changes are assumed to occur via thermal fluctuations and therefore do not involve a local material instability driven by the applied stress. A well known example of this approach is the so-called Activation Relaxation Technique $nouveau$ (ART$n$)~\cite{Barkema1996,Mousseau1998,Olsen2004} which has been applied to both two and three dimensional model glass systems~\cite{Rodney2009a,Rodney2009b,Kallel2010,Koziatek2013}. Starting from a well defined local minimum the ART$n$ method uses the local Hessian structure of the PEL to climb out of its current PEL valley and to eventually converge to a nearby saddle point region. The energy difference between the minimum and the saddle point gives the corresponding energy barrier of the associated LSE. When applied to model glass systems a wide range of energy barriers are obtained producing a distribution of energy barriers that appear to converge when several thousand such LSEs have been identified. For sufficiently relaxed glass configurations, the barrier energy distribution is found to be single peaked and analysis of the associated LSEs revealed their corresponding plastic strain increments to be uncorrelated with barrier energy~\cite{Rodney2009a}. Moreover, with respect to a particular loading geometry the plastic strains have equal numbers of positive and negative sign. However, upon application of the loading geometry strain, the distribution becomes slightly weighted towards positive strains with a large number of negative strain LSEs still occurring. The situation is quite different for the case of stress driven energy barriers which generally only yield LSEs that are positive in strain. Thus athermal stress driven simulations probe only a subset of possible LSEs --- those compatible with the chosen driving mode.

In this paper, the ART$n$ method will be applied to three-dimensional model LJ structural glass configurations to obtain a large number of LSEs and their associated barrier energies. The present work will only consider the case of zero external load. Using this data, the spatial location of the identified LSEs will be investigated in terms of local structural properties and a statistical analysis will be performed. The local quantities investigated will be atom type, local stress, local elastic constants, local pressure and Voronoi volume. In addition, the spatial nature of the structural changes and the number of atoms involved will be determined. All such information will also be correlated with the barrier energy. In sec.~\ref{SecMeth} the sample preparation procedure will be outlined and the local structural quantities to be investigated defined. Sec.~\ref{SecResults} contains the major results of the present work encompassing the ART$n$ results and the ensuing statistical analysis of the identified LSEs. In sec.~\ref{SecAV}, the spatial nature of a number of representative LSEs will be atomistically visualized and characterised. Finally, sec.~\ref{SecDiscussion} will discuss the results in terms of contemporary pictures of microscopic plastic deformation in BMGs.

\section{Methodology} \label{SecMeth}

\subsection{Sample preparation and ART$n$ calculations} \label{SecAtomistics}

Model glass samples have been prepared by molecular dynamics and statics using a 50/50 binary mixture of a LJ system given by,
\begin{equation}
V_{LJ}(r)=4\epsilon\left(\left(\frac {\sigma_{\alpha \beta}}{r}\right)^{12}-\left(\frac {\sigma_{\alpha \beta}}{r}\right)^6\right),
\end{equation}
where $\epsilon$ and $\sigma$ set the microscopic energy and length scale of the model material. The parameterization presently used is that of Wahnstr\"{o}m parametrization \cite{Wahnstrom1991}, which is shown in table \ref{Table1}. All simulation results are reported in LJ units where time is measured with respect to $\tau=\sqrt{m\sigma^2/\varepsilon}$ and temperature with respect to $\varepsilon/k_{\mathrm{b}}$.

\begin{table}[h]
\centering
\begin{tabular}{|l|l|c|c|c|} \hline
Parameter & Parameter Description & Value \\ \hline
$\varepsilon$ & Depth of the potential well & 1 \\ \hline
$\sigma_{11}$ & Diameter of atom type 1 & 1 \\ \hline
$\sigma_{12}$ & Cross-interaction diameter & 11/12 \\ \hline
$\sigma_{21}$ & Cross-interaction diameter & 11/12 \\ \hline
$\sigma_{22}$ & Diameter of atom type 2 & 5/6 \\ \hline
$r_c$ & Potential cut-off & 2.5$\sigma$ \\ \hline
$m_1$ & Mass of atom type 1 & 2 \\ \hline
$m_2$ & Mass of atom type 2 & 1 \\ \hline
\end{tabular}
\caption{Numerical values of the parameters of the LJ potential used in the present work \cite{Wahnstrom1991}, where $\sigma_{ij}$ represents the length-scale parameter between atoms of type $i$ and $j$.}
\label{Table1}
\end{table}

Four samples, each with 1728 atoms, have been prepared using different quench-rates ($\eta_1$ = $24.57/500$, $\eta_2$ = $24.57/5000$, $\eta_3$ = $24.57/50000$ and $\eta_4$ = $24.57/50000$). The sample preparation involves three steps, 1) Equilibration of the liquid state by NPT molecular dynamics at a temperature of $10000 \times k_b$ $[\varepsilon/k_b]$ and hydrostatic pressure of 8/160 $[\varepsilon/\sigma_{11}^3]$, 2) slow quenching of the sample from this well-equilibrated liquid state, which involves an incremental reduction in both temperature ($-198.0 \times k_b$ $[\varepsilon/k_b]$) and pressure (-0.158/160 $[\varepsilon/\sigma_{11}^3]$ ) by NPT molecular dynamics to form the disordered amorphous glass at a temperature of $100 \times k_b$ $[\varepsilon/k_b]$ and hydrostatic pressure of 0.1/160 $[\varepsilon/\sigma_{11}^3]$  3) Relaxation of the atomic coordinates to zero temperature and zero hydrostatic pressure by molecular statics using the Parrinello-Rahman method \cite{Parrinello1981}. For steps 1 and 2 the Parrinello-Rahman \cite{Parrinello1981} barostat has been used for pressure control and the Anderson-Hoover \cite{Anderson1985} thermostat has been used for temperature control.

The ART$n$ technique is then applied to the these samples. The ART$n$ identification of an LSE involves randomly choosing one atom and displacing it by a small distance ($<0.1\sigma$) from its equilibrium position --- the starting condition for ART$n$. The lowest eigen-value of the corresponding Hessian matrix is determined and the system is moved along the corresponding $3N$ dimensional eigen-vector until the eigenvalue of the Hessian becomes negative. This part of the algorithm is referred to as the Activation phase. At this point, the configuration has passed an inflection region of the PEL and enters the new phase of Relaxation. Here the configuration is moved in the direction of the eigenvector of the lowest eigen-value which is now negative, with a new eigen-value and eigen-vector begin calculated at each iteration. This is repeated until the dot-product of the total force of the configuration with the eigen-vector is zero. When this occurs, the configuration has reached a saddle-point which, following~\cite{Rodney2009a} is referred to as the Activated state.

To study the atomic scale environment of each identified LSE, an appropriate atomic weight for each atom is calculated via the displacement vectors between the initial state and either the activated or final state configuration. Presently the normalised weights (which sum to unity) are defined to be
\begin{equation}
w_{i}=\frac{\left|\Delta\textbf{R}_{i}\right|^{4}}{\sum_{i=1}^{N}\left|\Delta\textbf{R}_{i}\right|^{4}}. \label{EqWeights}
\end{equation}
This form was motivated by the standard definition of participation number of an eigenvector \cite{Bell1970}, which gives information about the number of elements contributing to the norm of the vector. Indeed,
\begin{equation}
\mathrm{PN}=\frac{1}{\sum_{i=1}^{N}w^{2}_{i}}, \label{EqPN}
\end{equation}
gives the effective number of atoms involved in the identified LSE. 

Given a Local Atomic Quantity (LAQ) for each atom, the weighted average of the quantity, according to eqn.~\ref{EqWeights}, will give a representative value for the region occupied by a particular LSE. That is,
\begin{equation}
\mathrm{LAQ}_{\mathrm{LSE}}=\sum_{i=1}^{N}w_{i}\times\mathrm{LAQ}_{i}. \label{EqLAQ}
\end{equation}
With this quantity, an average over many LSEs can be made and compared to the unweighted LAQ average where all atoms equally contribute to the average. Scatter diagrams for the weighted LAQ with respect to barrier energy are also investigated to determine if any correlation exists between local structural features and the activation energy of an LSE. To find out any possible linear correlation of such plots, the Pearson coefficient is used. For $n$ data points, this is given by
\begin {equation}
\mathrm{PC}=\frac{\sum_{i=1}^n(X_i-\bar{X})(Y_i-\bar{Y})}{\sqrt{\sum_{i=1}^n(X_i-\bar{X})^2}
\sqrt{\sum_{i=1}^n(Y_i-\bar{Y})^2}}, \label{EqPC}
\end {equation}
where ${X_i}$ and ${Y_i}$ are the data-sets, and $\bar{X}$ and $\bar{Y}$ are their respective arithmetic means. The value of PC ranges from $-1$ to $1$, where the values $-1$ and $1$ refer to a perfect linear correlation, and a value of zero indicates no {\em linear} correlation.

\subsection{Local Atomic Quantities} \label{SecLAQ}

The local quantities (the LAQs) presently considered are
\begin{itemize}
\item volume, calculated via an atomic scale Voronoi tessellation via the voro++ package \cite{Rycroft2006}
\item energy
\item pressure
\item dilation elastic modulus (three times the local bulk modulus)
\item the five linearly independent Kelvin eigen-shear elastic moduli
\end{itemize}

Since a LJ pair potential is presently being used, the expressions for the local stress and elastic modulus are of a simple form, where respectively
\begin{equation}
\sigma^{\mu\nu}_{a}=\frac{1}{2V_{a}}\sum_{ij}V'\left(R_{ij}\right)\frac{R_{ij}^{\mu}R_{ij}^{\nu}}{R_{ij}}\Lambda_{a,ij}
\label{EqnLStress}
\end{equation}
and
\begin{eqnarray}
C^{\mu\nu\alpha\beta}_{a}&=&\frac{1}{2V_{a}}\sum_{ij}\left[V''\left(R_{ij}\right)-\frac{V'\left(R_{ij}\right)}{R_{ij}}\right]\times\frac{R_{ij}^{\mu}R_{ij}^{\nu}R_{ij}^{\alpha}R_{ij}^{\beta}}{R_{ij}^{2}}\Lambda_{a,ij} \nonumber \\
& &+\sigma^{\nu\beta}_{a}\delta_{\mu\alpha}+\sigma^{\nu\alpha}_{a}\delta_{\mu\beta}.
\label{EqnLElasticConstants}
\end{eqnarray}
In the above expressions, $\Lambda_{a,ij}$ represents the proportion of the $ij$th bond within with the volume of atom $a$. It is noted that bonds between two atoms, neither of which is atom $a$, may also contribute to these two local quantities. Eqns.~\ref{EqnLStress} and~\ref{EqnLElasticConstants} properly partition volume and therefore correctly take into account the contribution of each atomic bond~\cite{Lutsko1988,Cormier2001}. The local pressure is obtained by taking one-third the trace of the local shear stress tensor. To obtain the Kelvin elastic moduli~\cite{Thompson1856} from the fourth rank elastic stiffness tensor (eqn.~\ref{EqnLElasticConstants}), the usual Voigt elastic stiffness matrix is first constructed from which the Kelvin matrix (2nd rank tensor) is obtained via $C^{\mu\nu}_{K}=A^{\mu\nu}C^{\mu\nu}_{V}$. For an explicit form of $A^{\mu\nu}$ see ref.~\cite{Derlet2012}. The five linearly independent eigen-shear moduli are obtained by first projecting out the pure dilation distortions and then diagonalizing the resulting Kelvin matrix --- for more details, see refs.~\cite{Mayr2009,Derlet2012}.

In contrast to the popular Voigt notation, the Kelvin notation preserves the norm of the actual elastic stiffness tensor and hence their eigen-values and eigen-vectors have geometrical significance~\cite{Mavko2009}. Since the Kelvin matrix is a tensor, the eigen-values of the stiffness matrix can be computed to obtain the bulk modulus and the eigen-shear moduli. The invariance of these eigen-shear moduli with respect to coordinate systems and thereby their role as an intrinsic material property has been highlighted in~\cite{Carcione1994}.

\subsection{Natural Mode Analysis}

The natural modes of an $N$ atom configuration can be obtained via the solution to
\begin{equation}
\sum_{j\nu}\left(m_{i}\left[\omega_{n}\right]^2\delta_{ij}\delta^{\mu\nu}-\Delta_{ij}^{\mu\nu}\right)u_{j,n}^{\nu}=0 \label{EqnSec}
\end{equation}
where $\Delta_{ij}^{\mu\nu}$ is the translationally invariant dynamical matrix obtained from
\begin{equation}
\Delta_{ij}^{\mu\nu}=\sum_{a,a\ne i}H_{ia}^{\mu\nu}\delta_{ij}-H_{ij}^{\mu\nu}\left(1-\delta_{ij}\right). \label{EqnHessianForm}
\end{equation}
Here $H_{ij}^{\mu\nu}$ is the Hessian. In terms of the LJ interaction, $V(r)$, the Hessian may be written as
\begin{equation}
H_{ij}^{\mu\nu}=\left[V''\left(R_{ij}\right)-\frac{V'\left(R_{ij}\right)}{R_{ij}}\right]\frac{R_{ij}^{\mu}R_{ij}^{\nu}}{R_{ij}^{2}}+
\frac{V'\left(R_{ij}\right)}{R_{ij}}\delta^{\mu\nu}. \label{EqnHessianInteraction}
\end{equation}
In eqn.~\ref{EqnSec}, $m_{i}$ is the atomic mass of the $i$-th atom, and $u_{j,n}^{\nu}$ is the eigen-vector associated with the eigen-frequency $\omega_{n}$ of the $n$th natural mode. The number of atoms participating in a particular eigenstate, $u_{j,n}^{\nu}$, may be obtained via the participation number \cite{Bell1970}
\begin{equation}
\mathrm{PN}_{n}=\left[\sum_{i}\left|\vec{u}_{i,n}\right|^{4}\right]^{-1},
\label{EqnVibPN}
\end{equation}
where $\vec{u}_{i,n}$ is the three-dimensional polarisation vector of atom $i$ coming from the eigen-vector of eigen-frequency $\omega_{n}$. Assuming a normalised eigen-vector, $\mathrm{PN}_{n}$ will range between unity (when the eigenstate is concentrated on just one atom) and $N$ (when the eigenstate is distributed evenly over the entire sample).

\section{Results} \label{SecResults}

\subsection{ART$n$} \label{SecART}

In total 1347 unique activated states where identified using the ART$n$ method. To verify that each activated state was directly connected to the initial atomic configuration, the activated configuration was perturbed in a direction towards the initial atomic state configuration and allowed to relax. If the resulting structure differed from the initial structure the LSE was discarded from the data-set, as was done in ref.~\cite{Rodney2009a}. In a similar way, the final state could be determined by perturbing the activated configuration in a direction away from the initial atomic state configuration and allowed to relax. In total 954 initial/activated/final state atomic configurations where obtained for subsequent analysis. 

\begin{figure}
\centering
\includegraphics[clip,scale=0.45]{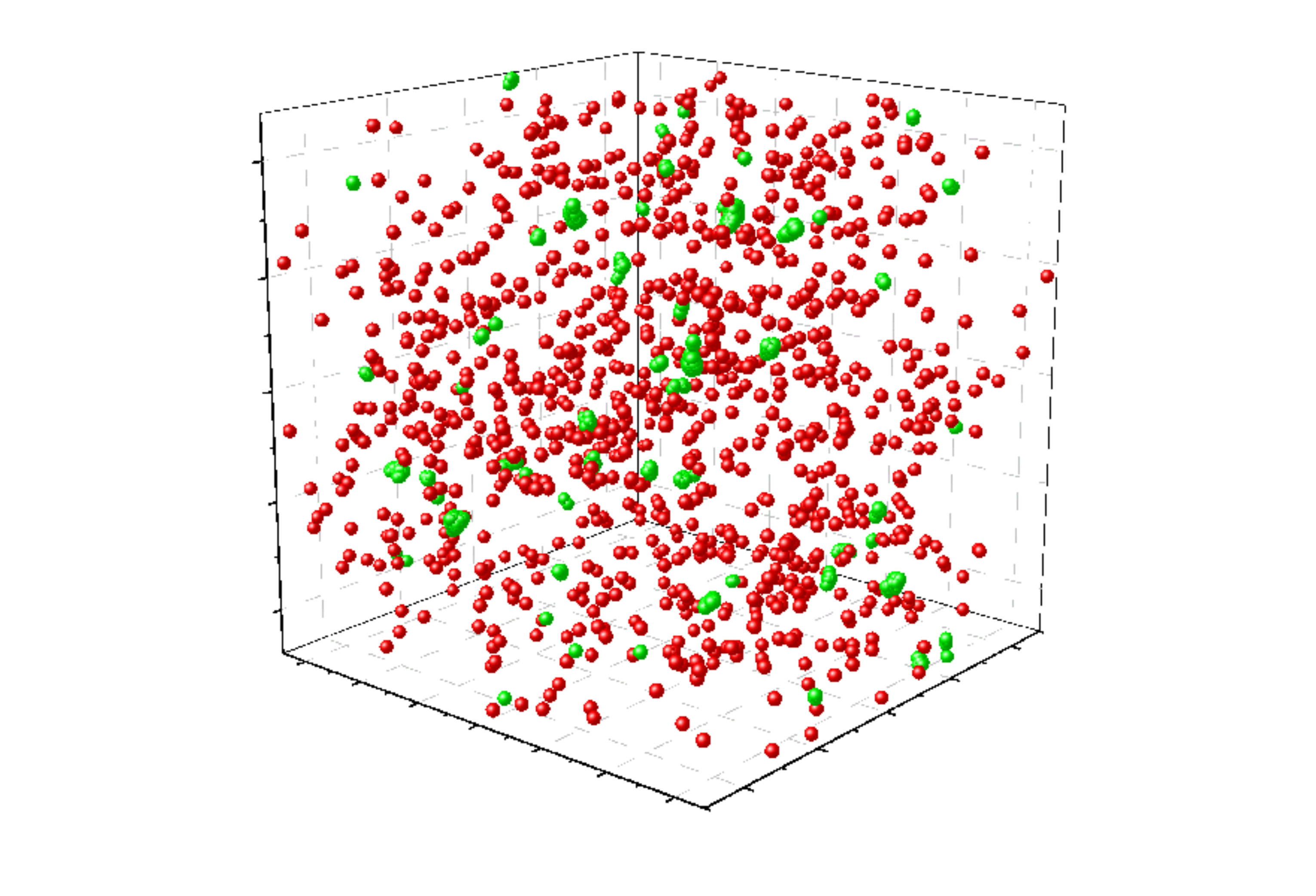}
\caption{\label{FigCOP} Red balls represent the centre-of-positions of all identified local structural excitations and the green balls represent regions containing free volume within the simulation cell of the model glass}
\end{figure}

To determine the spatial location of a particular LSE, the centre-of-position of those saddle-point atoms displaced relative to the initial configuration by more than $0.1\sigma$, was calculated. Fig.~\ref{FigCOP} displays these positions within a boundary box defining the three dimensional periodic simulation cell. A general inspection of their spatial distribution reveals some heterogeneity and, in particular, regions where many LSEs are similarly located. A more detailed inspection of such LSEs (as in a manner described in sec.~\ref{SecAV}) reveals them to be quite different in spatial extent and barrier energy, despite some LSEs having their centre-of-position almost coincident. The goal of the proceeding section will be to see if any local structural feature correlates with this observed heterogeneity.

\begin{figure}
\centering
\includegraphics[clip,scale=0.75]{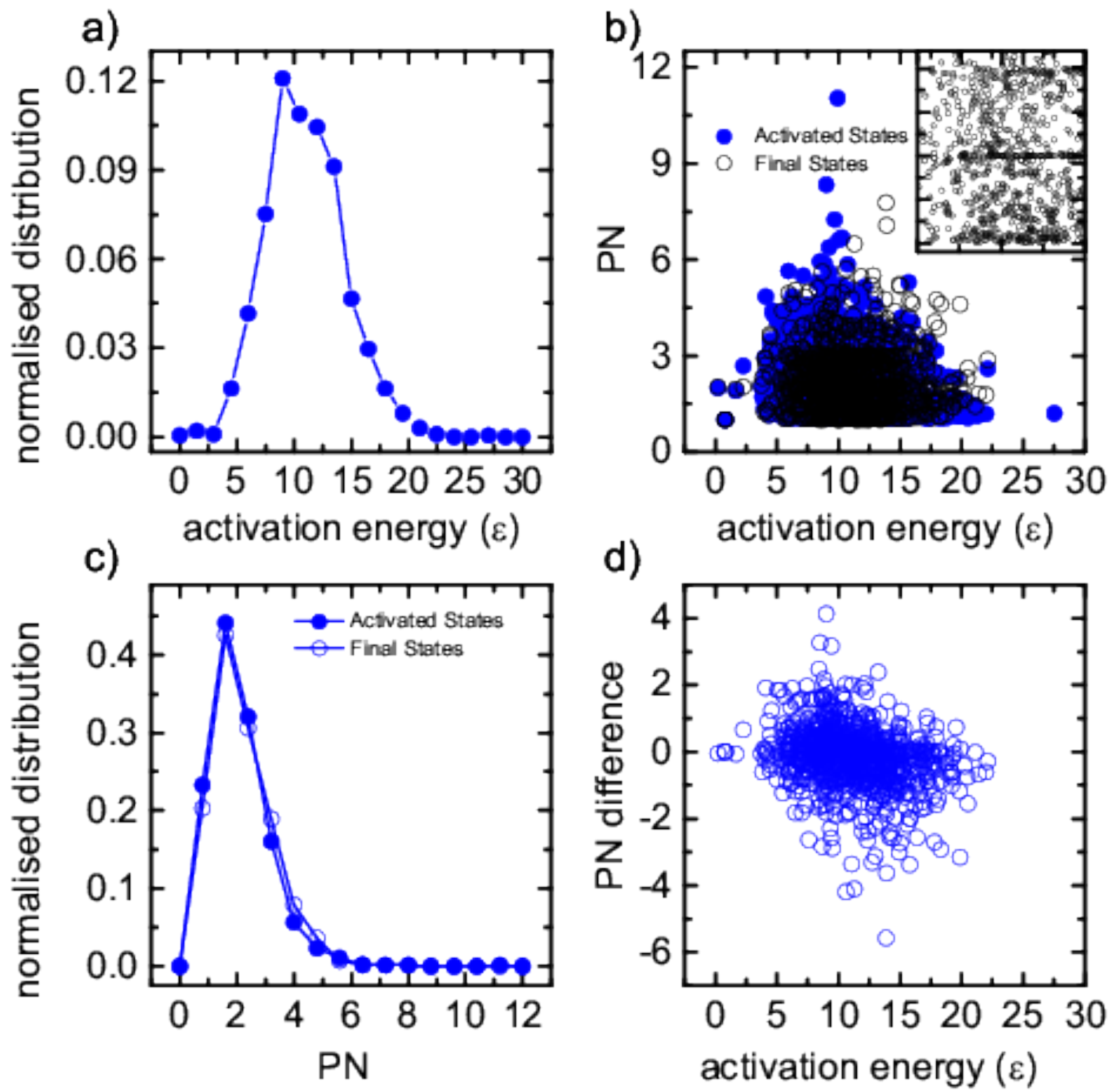}
\caption{\label{FigBED} a) Normalised distribution of activation energies of LSEs in a 3D model glass system. b) Scatter plot of participation number (PN) of activated and final state configurations versus activation energy where the inset is a blow up of the final state configuration participation numbers. c) Histogram of participation number and d) scatter plot of participation number difference between the connected activated and final states as a function of activation energy.}
\end{figure}

Fig.~\ref{FigBED}a displays the distribution of barrier energies obtained from the 954 identified activated states for the sample with the slowest quench rate. In agreement with refs.~\cite{Rodney2009a,Rodney2009b,Kallel2010,Koziatek2013}, the distribution peaks at a non-zero barrier energy and appears to approach zero for small enough barrier energies. The barrier energy scale is comparable to that seen in a previous fully three dimensional ART$n$ simulations using the same LJ potential parametrisation~\cite{Koziatek2013}. Inspection of the participation number in fig.~\ref{FigBED}b reveals that the effective number of atoms involved in the LSE is typically less than ten and that the LSEs with barrier energies in the lowest and highest regime involve only a few atoms. Close inspection of the final state participation number (see inset of fig.~\ref{FigBED}b) reveals a clustering at integer values of the participation number. This does not occur for the activated state configuration demonstrating that whilst the non-affine displacement field associated with the activated configuration is dispersed over many atoms, the non-affine field of the final state can be localized on a discrete number of atoms. Further inspection of all LSEs revealed that 81 initial/activated/final state atomic configurations had the feature that the final state configuration had an identical total energy to the initial state. Detailed inspection of these LSEs revealed that in these 81 cases, the final state involved a permutation of nearby atoms of the same type, where the activated configuration involved a closed loop of displaced neighbouring atoms (see fig.~\ref{FigVis}a for an example). Thus the final state is identical to the initial state and when not included in fig.~\ref{FigBED}b the discrete participation numbers vanish. Since such LSEs cannot produce any strain these were also removed from the data-set used in the LAQ analysis of proceeding section.

Fig.~\ref{FigBED}c now displays a histogram of the participation numbers indicating that the LSEs identified by ART$n$ generally involve somewhere between one and several atoms. Thus on average there is little difference in the number of participating atoms between a activated and final state. On the other hand, fig.~\ref{FigBED}d displays the difference in participation number between connected activated and final state configurations showing that the number can either decrease or increase by several atoms. On average, however, the change in the participation number is close to zero indicating no strong bias to whether the final state contains more or less participating atoms than the activated state, a result confirming fig.~\ref{FigBED}c.

\subsection{Statistical analysis of LAQs} \label{SecSA}

\begin{figure}
\centering
\includegraphics[scale=0.75]{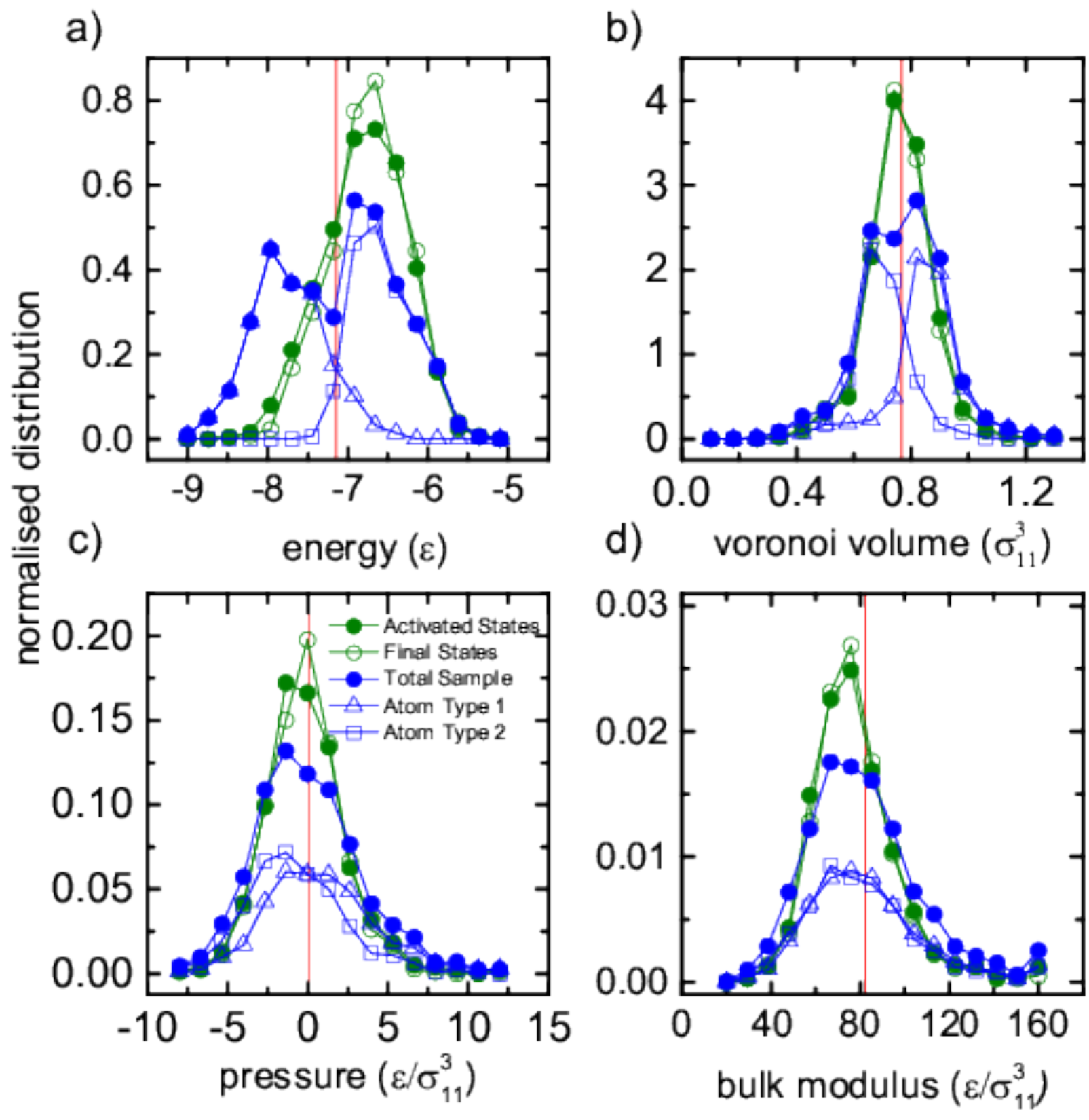}
\caption{\label{FigLAQ} Normalized distribution of local a) energy b) volume c) pressure d) bulk modulus. The red vertical lines represent the corresponding mean value derived from the total sample.}
\end{figure}

Fig.~\ref{FigLAQ} displays the normalised histograms of a number of LSE averaged LAQs with respect to the activated and final state configurations, also shown are the equivalent unweighted histograms derived from the total sample. The two types of distributions will be referred to as LSE weighted and unweighted distributions of the LAQ. For the unweighted distributions the (atom-type resolved) partial histograms are also shown.

Fig~\ref{FigLAQ}a represents the normalised distribution of potential energy of atoms. The double peak structure of the total unweighted distribution is clearly seen to arise from the single peaked distributions of each atomic type. The LSE weighted distributions, on the other hand, do not exhibit a double peak structure, with the single observed peak coinciding with the unweighted partial distribution of atoms of type 2. This result suggests that the atoms involved in an LSE are often of type 2. Fig.~\ref{FigLAQ}b now shows the corresponding normalised distribution for the Voronoi volume. Inspection of the unweighted curves show that atoms of type 2 have lower volume compared to atoms of type 1, a feature that is expected given the nature of the Wahnstr\"{o}m parametrization (see table~\ref{Table1} and ref.~\cite{Derlet2012}). The distinct double peak feature is however absent for the LSE weighted distributions, both the activated and final state curves showing a single peaked structure approximately centered between the unweighted partial distributions. Closer inspection does however reveal a slight bias to the lower volumes of the type 2 atoms. Figs.~\ref{FigLAQ}c and d show respectively the distributions for both local pressure and local bulk modulus. For both LAQs, the unweighted total distribution show single peaked curves that are slightly asymmetric due to slightly different contributions from the two atom types. That atoms of type 2 are involved in the LSEs is also reflected here because the LSE weighted curves are more biased toward the partial unweighted curves of type 2 atoms.

\begin{figure}
\centering
\includegraphics[scale=0.75]{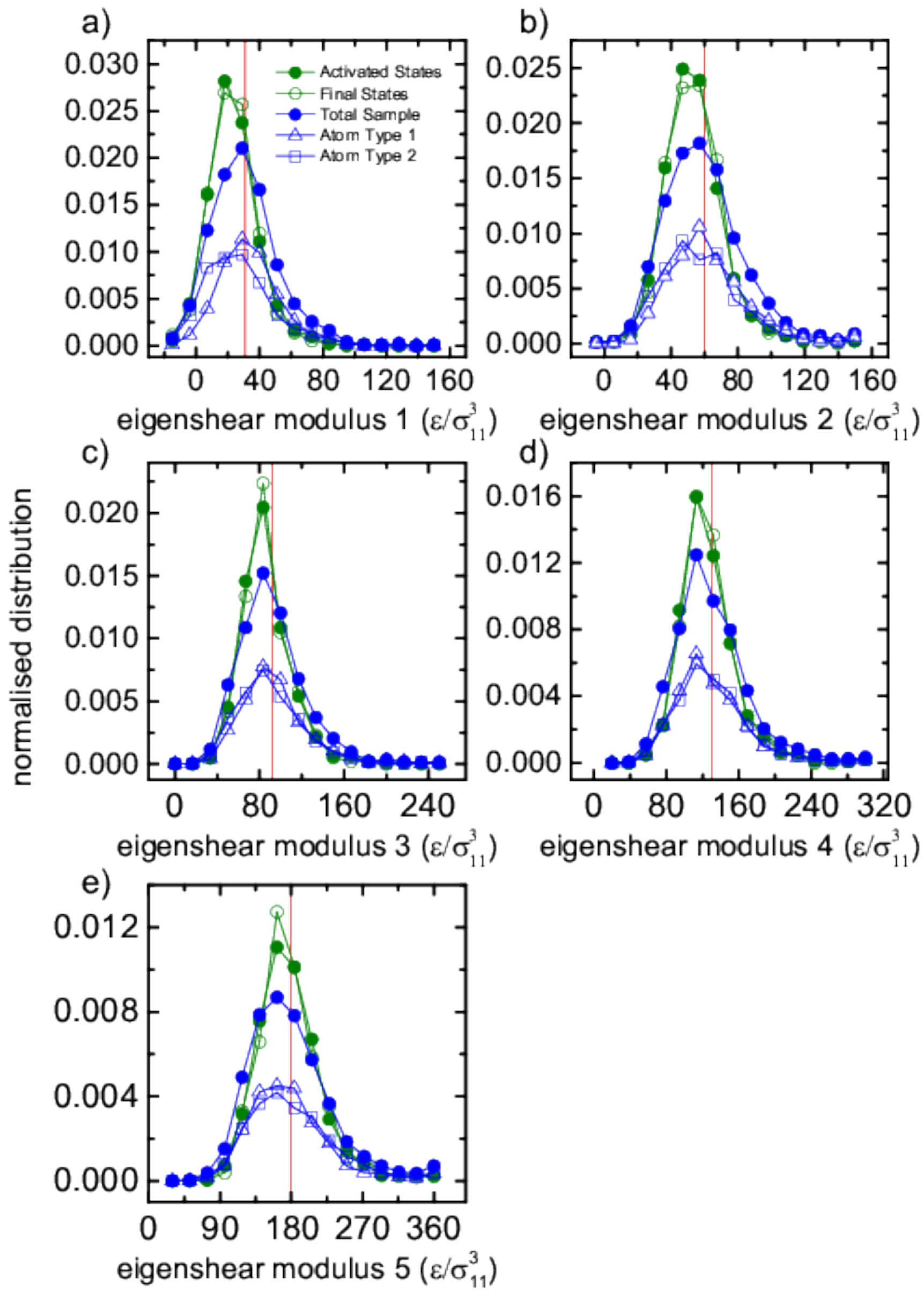}
\caption{\label{FigLAQES} Normalised distribution of five local eigenshear moduli where a) to e) represents the lowest to highest values. The red vertical lines represent the corresponding mean value derived from the total sample.}
\end{figure}

Fig~\ref{FigLAQES}a shows the distribution of Kelvin eigen-shear moduli for the activated and final relaxed states. Again, the unweighted total and partial distributions are shown for comparison. It is noted that for each atom, the local Voigt matrix is first constructed via eqn.~\ref{EqnLElasticConstants}, from which the local Kelvin matrix is built and is then diagonalized (after the dilation components are projected out) to obtain five eigen-shear moduli. These are then ordered and each order is binned separately to produce the five panels of fig.~\ref{FigLAQES}. As seen in ref.~\cite{Derlet2012}, the left tail of the distribution of lowest eigen-shear moduli extends into the negative moduli domain. That some atoms have a local distortion characterised by a negative modulus does not entail a local material instability since their calculation involves only those neighbouring atoms with a direct interaction and not the stabilising effect of the more distant surrounding matrix. Such low or negative eigen-shear moduli do however indicate the presence of local shear distortions that are soft. Inspection of the unweighted total and partial LAQ single peak distributions reveal that atoms of type 2 are slightly biased towards regions of softer moduli. The weighted LAQ single peak distribution follows this bias, confirming that atoms of type 2 are often involved in an LSE.

\begin{figure}
\centering
\includegraphics[scale=0.75]{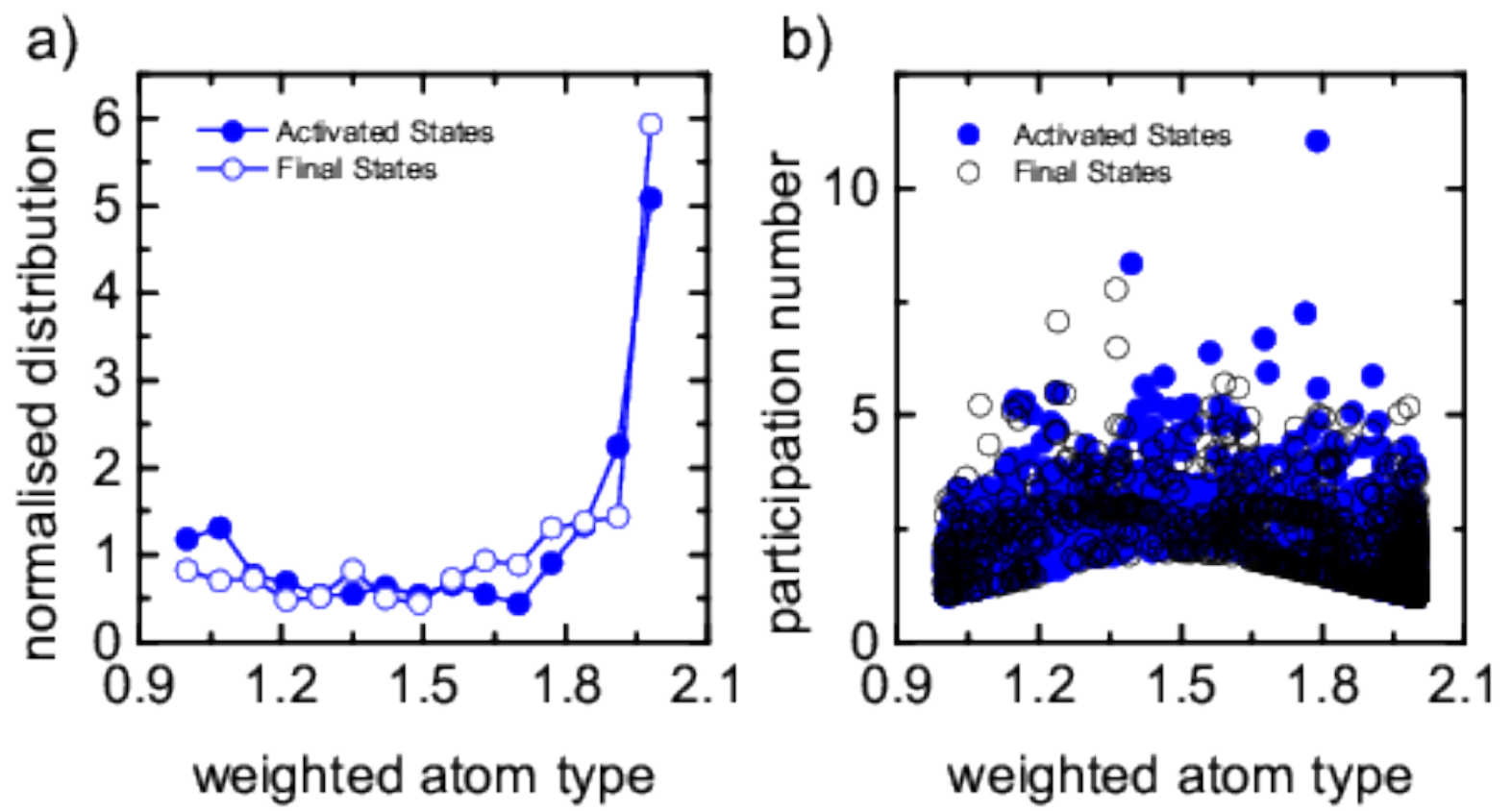}
\caption{\label{FigType} a) Normalised distribution of LSE weighted atom types for the activated and final state configurations and b) Scatter plot of LSE weighted atom type against participation number.}
\end{figure}

To gain more direct information on the type of atom involved, normalised atom type LSE weighted distributions are generated. Fig.~\ref{FigType}a displays these for both the activated states and the final states. Both distributions peak at the average atom-type of 1 and 2, and for intermediate values are essentially a flat distribution with a slight bias towards LSEs with higher average atom-type. This suggests that the most probable specific chemical composition will consist of only type 2 atoms, however on average an LSE will contain a mixture of both atom types with a bias towards atoms of type 2. Inspection of those LSEs containing only atoms of type 2 reveals them to consist of only one or two atoms. On the other hand, LSEs containing a mixture of both atoms types can consist of many more atoms. This is demonstrated in fig.~\ref{FigType}b which shows the scatter plot of participation number with LSE weighted atom type. Here there is a clustering of small LSEs around atom type 2, but many larger LSEs consisting of both type 1 and type 2 atoms are evident. The lower boundary in this figure reflects the fact that for an average atom type of 1.5 at least two atoms must be involved.

From the perspective of thermal activation the rate of occurrence for a particular LSE is given by $\nu_{0}\exp(-E_{0}/k_{\mathrm{b}}T)$ where $\nu_{0}$ is the so-called pre-factor (rate of attempt) and $E_{0}$ is the barrier energy of the LSE. Therefore it will be those LSEs with the lowest barrier energy that will most likely occur. In the work of Koziatek {\em et al} \cite{Koziatek2013} this was demonstrated within harmonic transition state theory, where the corresponding pre-factor, $\nu$, was also calculated, showing that despite a wide range of pre-factor values (spanning many orders of magnitude) those identified LSE with lowest barrier energy generally exhibited the highest rate of occurrences. It is therefore of interest to investigate the correlation between barrier energy and the structural environment as presently defined by the LAQs of sec.~\ref{SecLAQ}.

\begin{figure}
\centering
\includegraphics[scale=0.75]{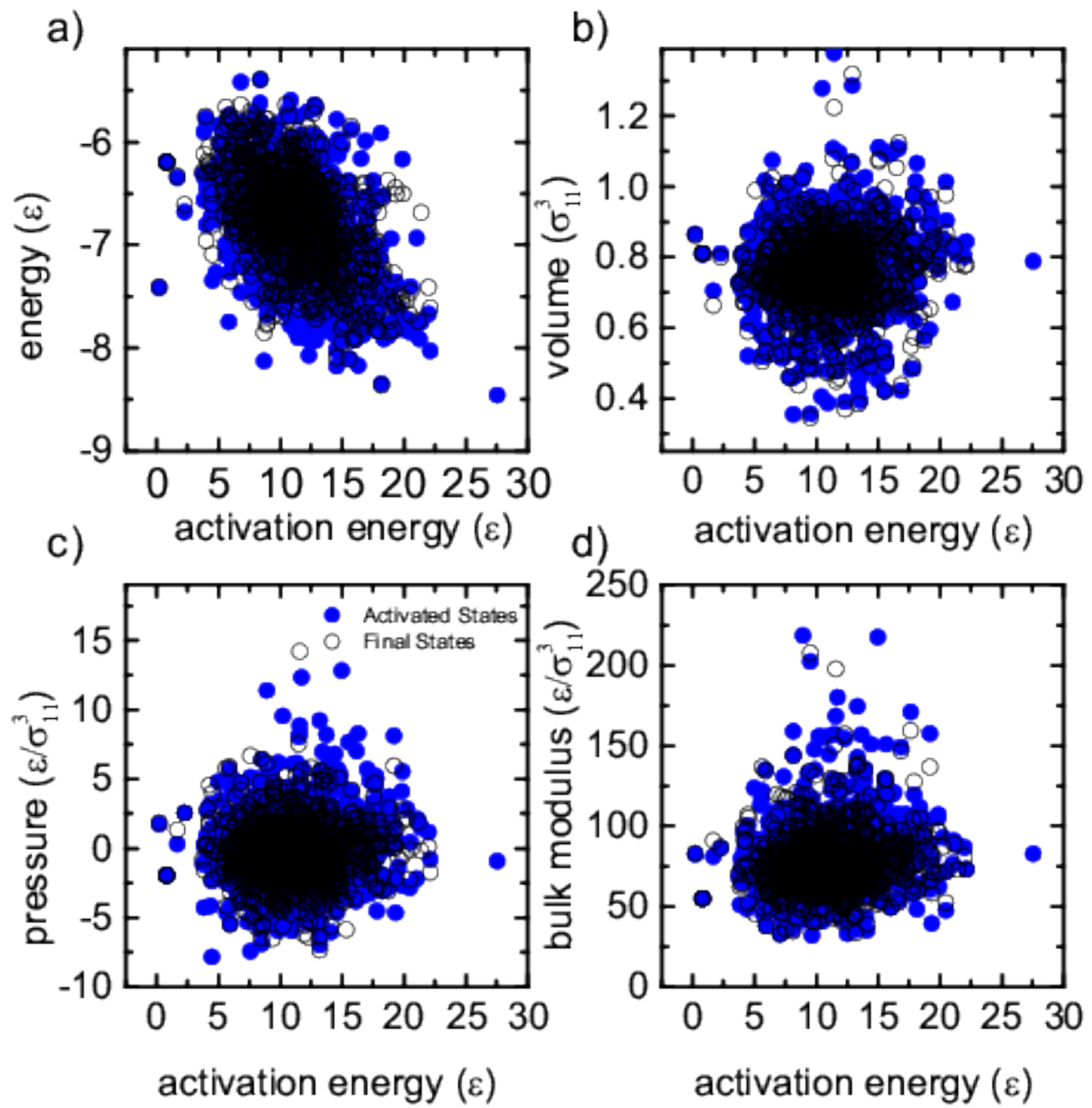}
\caption{\label{FigLAQScatter} Scatter plots of LSE weighted local a) energy b) volume c) pressure and d) bulk modulus against activation energies.}
\end{figure}

\begin{figure}
\centering
\includegraphics[scale=0.75]{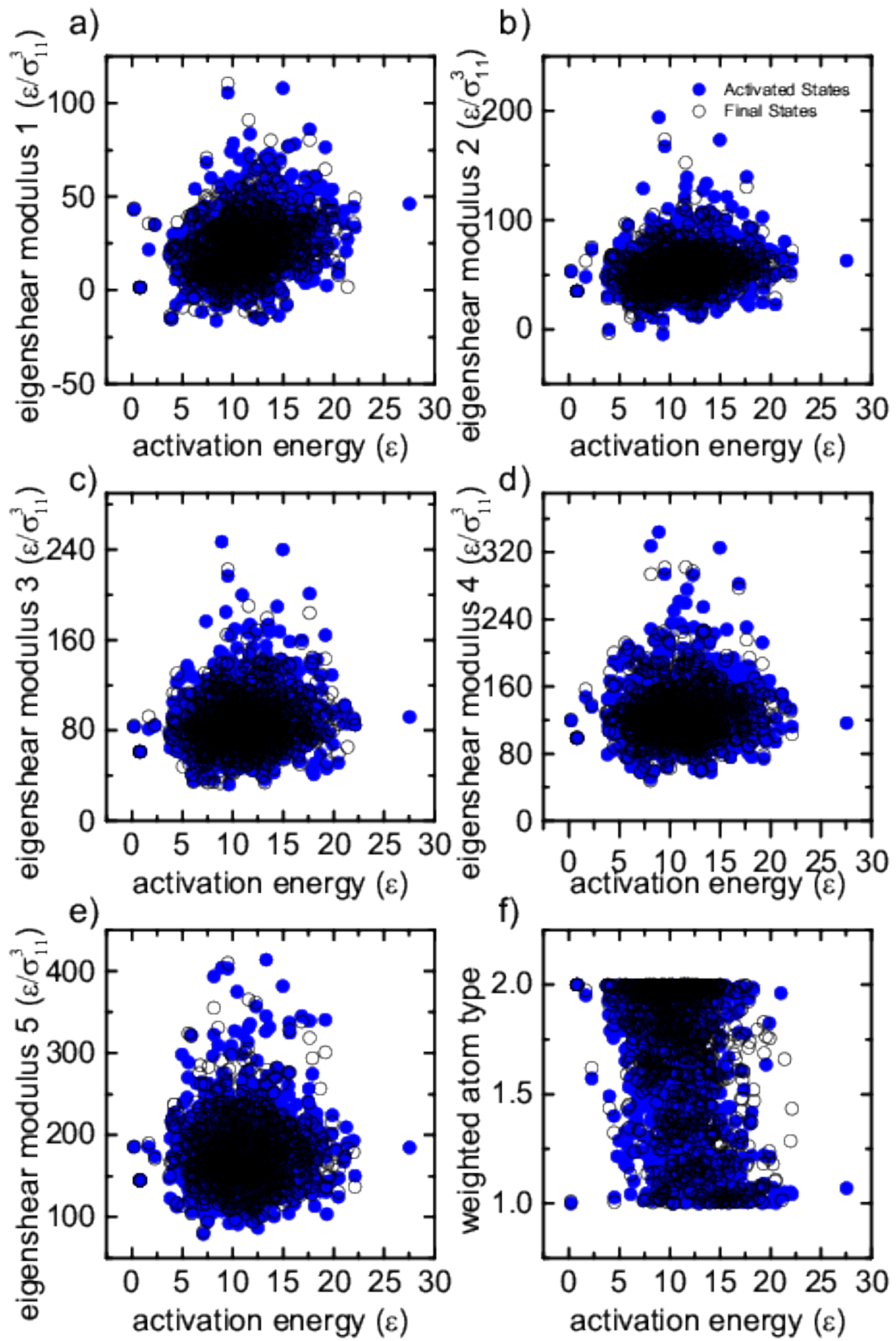}
\caption{\label{FigLAQScatterES} Scatter plots of LSE weighted local eigenshear moduli against activation energies from a) lowest to e) highest value. f) Scatter plot of atom type and activation energy}
\end{figure}

Fig.~\ref{FigLAQScatter} displays scatter plots of the LSE weighted LAQs with respect to their corresponding barrier energy. Data is shown for local cohesive energy, volume, pressure and bulk modulus. Inspection of these figures reveal strong scatter and no strong linear correlation. This is evidenced by the associated Pearson correlation coefficients shown in tab.~\ref{Table2}. The exception to this trend is that of local cohesive energy, which has a Pearson correlation coefficient equal to approximately -0.5, indicating a non-negligible linear correlation with the barrier energy. Fig.~\ref{FigLAQScatterES} shows similar scatter plots for the five Kelvin eigen-shear moduli. Little correlation is again evident, although the Pearson correlation coefficient for the lowest eigen-shear modulus is approximately 0.25 with the value progressively decreasing with increasing eigen-shear number --- see tab.~\ref{Table2} which lists these coefficients for all five moduli. What causes these weak correlations? A closer inspection demonstrates that the origin is again atom type, where those LSE with predominantly atoms of type 2 are not only more common, but they also appear to correspond to lower barrier energies. This is directly seen in fig.~\ref{FigLAQScatterES}f which is a scatter plot of atom type and barrier energy. This plot shows again that there is a bias towards LSEs with a majority of type two atoms with the corresponding barrier energy being on average slightly lower than the barrier energies associated with LSEs dominated by atoms of type one. The scatter associated with this trend is however large, with both types of LSEs having a spread in barrier energy comparable to the domain of the distribution shown in fig.~\ref{FigBED}a.

\begin{table}
\centering
\begin{tabular}{|l|c|c|c|c|c|} \hline
Quantity & Activated State PCC & Final Relaxed State PCC \\ \hline
 BM & 0.1593 & 0.1039 \\ \hline
 E & -0.4793 & -0.4897 \\ \hline
 P & 0.1264 & 0.0084 \\ \hline
V & 0.0821 & 0.0522  \\ \hline
ES1 & 0.2850 & 0.2473 \\ \hline
ES2 & 0.2170 & 0.1703 \\ \hline
ES3 & 0.1611 & 0.1343 \\ \hline
ES4 & 0.1051 & 0.0461 \\ \hline
ES5 & 0.0506 & 0.0154 \\ \hline
\end{tabular}
\caption{\label{Table2} Pearson Correlation Coefficients (PCC) representing the correlation between local atomic quantity and activation energy} \end{table}
\subsection{Correlation with Vibrational Modes} \label{SecVib}

\begin{figure}
\centering
\includegraphics[scale=0.75]{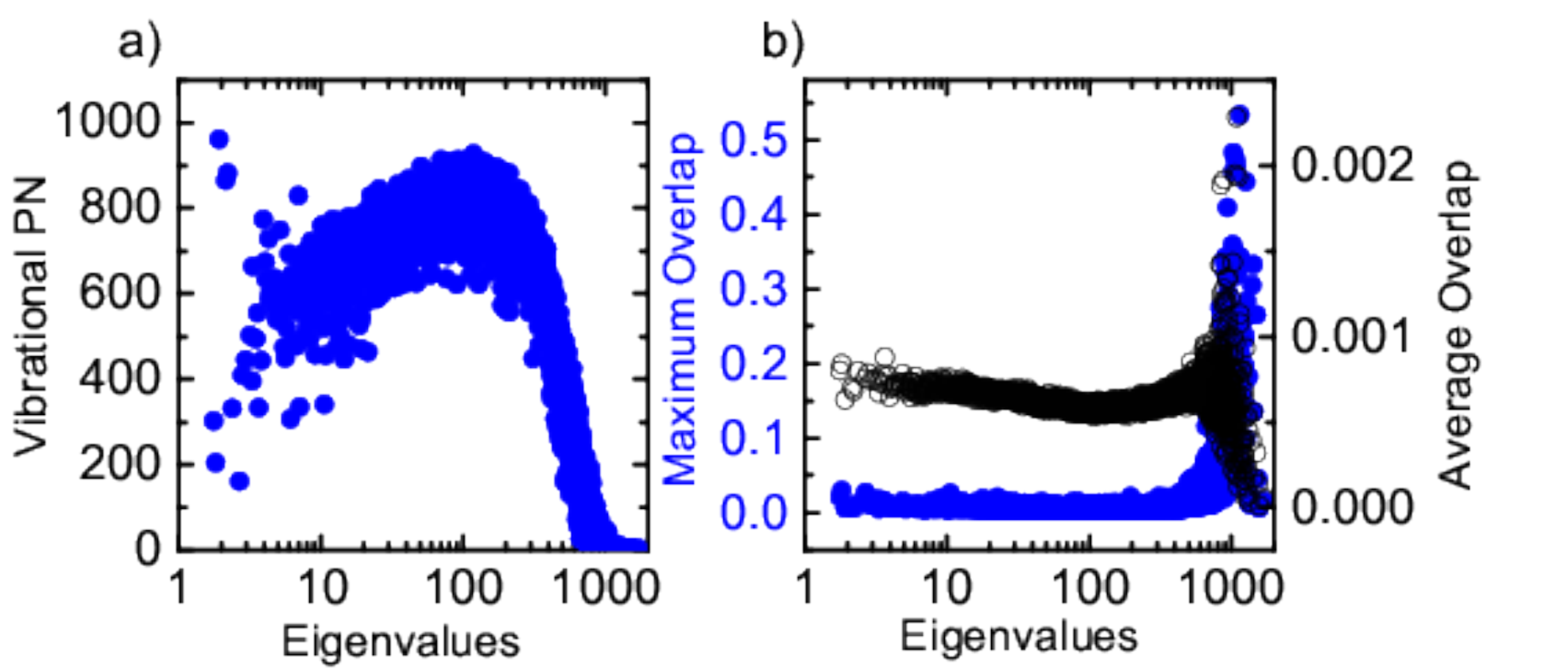}
\caption{\label{FigOverlap} a) Plot of vibrational participation numbers against eigenvalues b) Plot of maximum and average overlap per eigenvalue.}
\end{figure}

Fig.~\ref{FigOverlap}a shows the plot of the ``vibrational participation numbers'' (eqn.~\ref{EqnVibPN}) against eigenvalue number corresponding to each eigenstate, where the vibrational participation number represents the effective number of atoms participating in a vibrational eigen-mode. Here the eigenvalues have been sorted from smallest to largest and therefore the horizontal axis is proportional to increasing vibrational frequency. Such data has been calculated before~\cite{Schober1991,Mazzacurati1996,Schober1996,Derlet2012}, demonstrating that at low frequencies the eigen-modes are strongly heterogeneous indicating quasi-localized mode behaviour that is believed to underly the well known Boson peak phenomenon of disordered systems~\cite{Gurevich2003,Gurevich2005,Parshin2007,Shintani2008,Monaco2009a}. At the highest frequencies the participation number again drops reflecting a localization that is understood within the framework of Anderson localization~\cite{Anderson1958,Garber2001,Huang2009,Allen1999}. To investigate any correlation between the existence of such low-frequency quasi-localized and high-frequency localized modes, the overlap between the vibrational eigenstate and that of the LSE is determined. This is presently done by calculating the scalar product of the atomic weights (eqn.~\ref{EqWeights}) with the eigenvector magnitude-squared, $\left|\vec{u}_{i,n}\right|^{2}$. Fig.~\ref{FigOverlap}b displays both the average overlap and the maximum overlap of all identified LSEs with each vibrational eigen-mode.

The figure shows that there exists, on average, little overlap over the entire frequency range. In the low frequency regime, the average overlap is a well defined statistical quantity indicating that irrespective of the nature of the quasi-localized mode, the spatial location and extension of the identified LSEs are similar for different modes. This is also reflected in the maximum overlap which varies little with eigenstate, and is also a small quantity. At higher frequencies the situation is somewhat different in that there is much more scatter in the average value and that the maximum value. This however does not indicate any important correlation since the small average and large maximum values are more likely to indicate the scenario that statistically there will generally be one or some LSEs that strongly overlap with a one or some well-localized high frequency eigenstates. This does not occur at the low-frequency quasi-localized eigenstates since these are more extended involving several tens to hundreds of atoms (see fig.~13 of ref.~\cite{Derlet2012}). Thus the current analysis reveals little correlation between the spatial extent of the vibrational modes and the location of the identified LSEs.

\subsection{Effect of Quench Rates}

\begin{figure}
\centering
\includegraphics[scale=0.75]{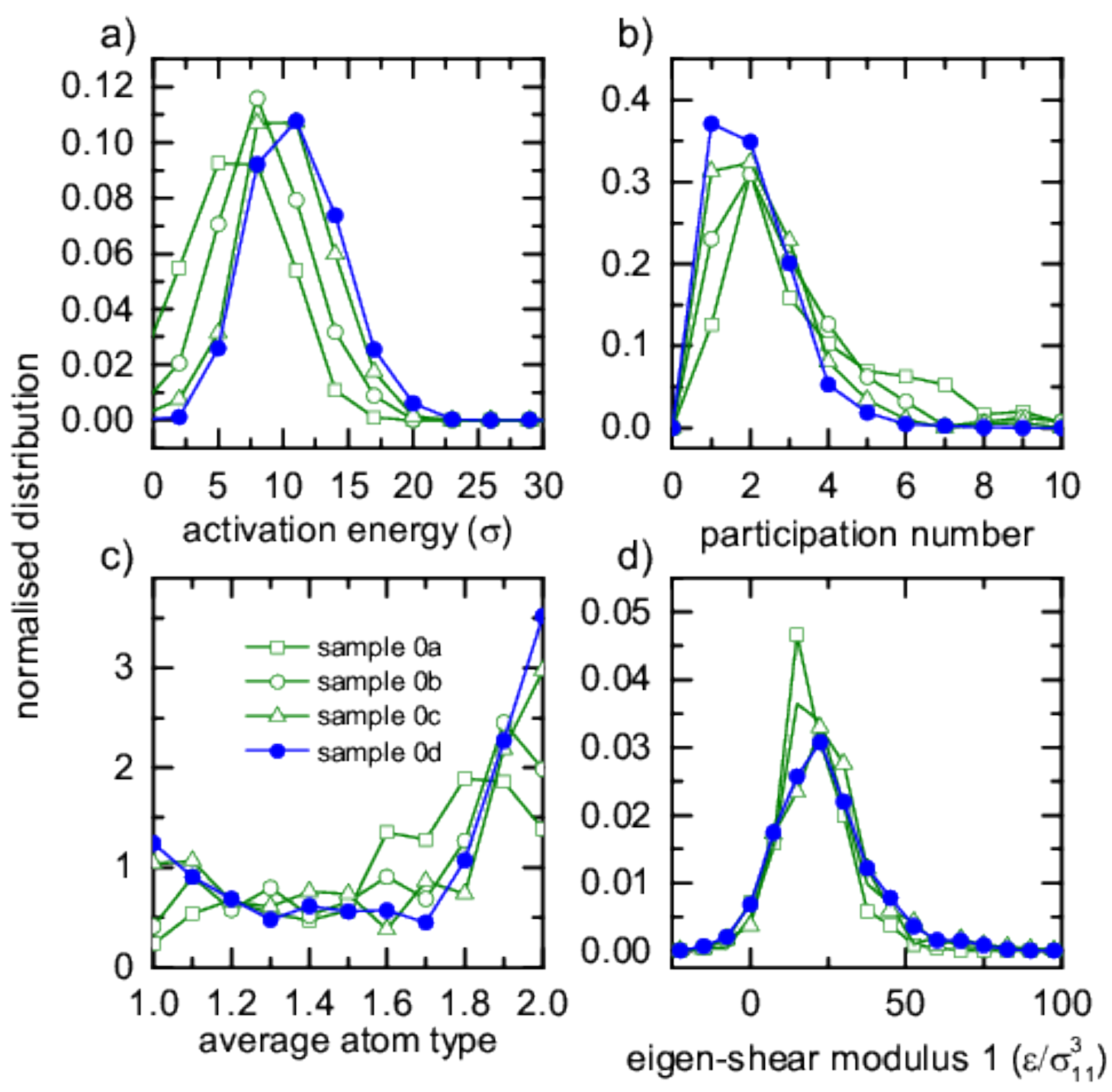}
\caption{\label{FigQR} Panels showing the effect of different quench rates in which the fastest quench rate is sample0a and the slowest is sample0d. a) Shows histograms of the activation energy, b) the participation number, c) the atom type and d) the lowest Kelvin eigenshear distribution.}
\end{figure}

The ART$n$ simulations have been also performed on glass samples prepared via faster quench rates (sample 0a with $\eta_1$, sample 0b with $\eta_2$, sample 0c with $\eta_3$), and a comparative study has been done to determine if the results of sec.~\ref{SecResults} depend on the quench rate. Fig.~\ref{FigQR}a displays the resulting activation energy distributions, showing that for the more rapidly quenched systems the peak of the distribution shifts to lower activation energies and that close to the zero activation energy limit, the distribution does not limit to zero (a result also found in the work of Rodney and Schuh~\cite{Rodney2009a,Rodney2009b}). Fig.~\ref{FigQR}b now shows the participation number distribution (eqn.~\ref{EqPN}) revealing that with increasing quench rate there is a slight shift to a larger number of atoms being involved in the LSEs and that many of these involve a more mixed number of atom type --- see fig.~\ref{FigQR}c which displays the average atom type distribution of the LSEs. These results tend to suggest that the more rapidly the model system is quenched the more shallow the local potential minimum is of each resulting 0K atomic configuration, and that the corresponding LSEs are somewhat larger involving both types (sizes) of atom. In general however the weak correlation with local structural features seen in sec.~\ref{SecResults} is insensitive to quench rate. For example, fig.~\ref{FigQR}d displays the LAQ weighted lowest Kelvin eigen-shear distribution as a function of quench rate, which displays only a slight shift towards lower elastic stiffness moduli.

\section{Atomic Visualisation} \label{SecAV}

\begin{figure}
\centering
\includegraphics[scale=0.75]{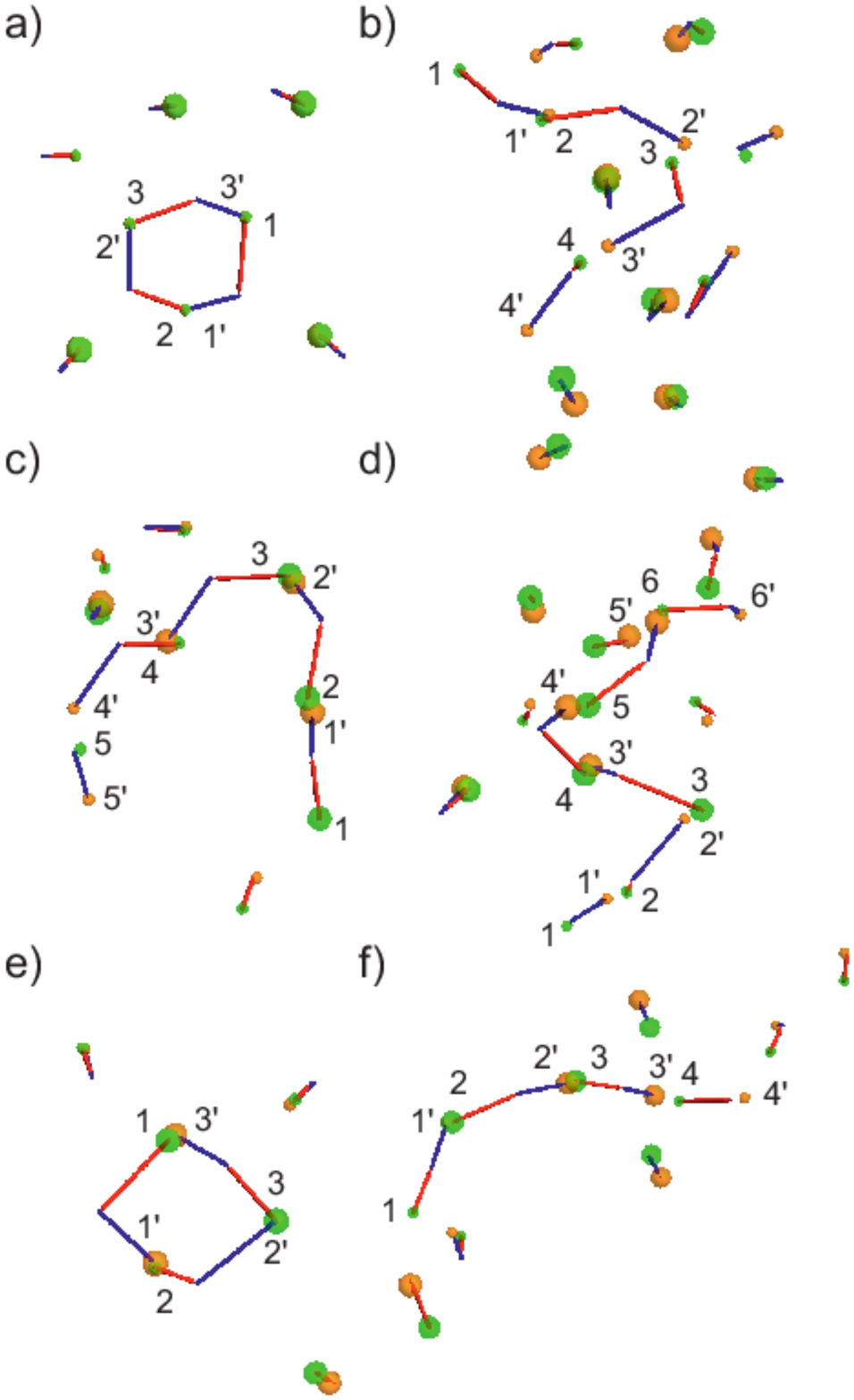}
\caption{\label{FigVis} Six examples of local structural excitations identified by ART$n$. In each case the initial atomic positions are visualized by green balls and the final by orange balls. The atomic displacements from the initial to activated and activated to final states are visualized by respectively red and blue arrows.}
\end{figure}

In this section six identified LSEs are atomistically visualized. These examples have been chosen since they represent typical features seen in all LSEs and are shown in fig~\ref{FigVis}. In all of the examples, the initial and final atomic positions are represented respectively by green and orange spheres, where the red arrows represent the displacement from the initial to activated position and the blue arrows the displacements from the activated to final position. Only those atoms are shown which are displaced by more than $0.2\sigma$, either between the initial and activated or the activated, or final configuration. The large spheres represent atoms of type 1 and the small spheres atoms of type 2. Generally, the visualised atoms may be classified into two groups, those central atoms that involve significant and irreversible displacement and those atoms that accommodate this activity either via reversible elastic or plastic activity. In all figures, the first class of initial atom positions are numbered, with the dashed-corresponding-number labeling their final position. 

Fig.~\ref{FigVis}a represents an LSE with an activation energy of $11.98\epsilon$ involving 8 atoms. In this LSE, the central atomic structure forms a symmetrical ring like (or closed chain-like) structure ($1\rightarrow1':2\rightarrow2':3\rightarrow3'$) consisting mainly of smaller atoms (of type 2). Surrounding this plastic inner structure, there are mainly larger atoms (of type 2) which move back and forth during the initial to activated state and then from activated to final state transition respectively. This is an example of elastic accommodation mechanism around the inner ring-like plastic rearrangement. This is an example of an LSE that results in a final configuration identical to the initial configuration apart from a permutation of three labels. It is such LSEs, numbering 81, that have been removed from the data-set studied in sec.~\ref{SecLAQ}.

Fig~\ref{FigVis}b represents an LSE with an activation energy of $6.65\epsilon$ involving 17 atoms. It shows an extended chain-like atomic motion with the sequence being specified by ($1\rightarrow1':2\rightarrow2':3\rightarrow3':4\rightarrow4'$). Although smaller atoms (of type 2) are involved in the formation of the chain, there are a relatively large number of large atoms (of type 1) which are responsible for accommodating this structural excitation. There is evidence of both elastic and plastic accommodation by the surrounding atoms, which is clearly seen by atoms moving back and forth as well as atoms moving irreversibly in the region surrounding the inner chain like formation. An observation that is inferred out of this LSE (and further confirmed in the subsequent descriptions of LSEs) is that, excitations which involve a higher number of atoms in the chain-like structure also involve a proportionally higher number of atomics in the surrounding accommodation mechanism. Fig~\ref{FigVis}c represents an LSE with an activation energy of $10.35\epsilon$ involving 9 atoms. It also shows a chain-like atomic reconfiguration, now of a strongly curved extension. The sequence is specified by ($1\rightarrow1':2\rightarrow2':3\rightarrow3':4\rightarrow4':5\rightarrow5'$). Here both sized atoms are involved in the re-configuration. There are two bigger atoms and one smaller atom surrounding this chain which undergo elastic displacement.

One of the most spatially extended LSEs identified by ART$n$ is shown in fig.~\ref{FigVis}d. This LSE has an activation energy of $16.01\epsilon$ and involves 17 atoms, with the reconfiguration sequence being ($1\rightarrow1':2\rightarrow2':3\rightarrow3':4\rightarrow4':5\rightarrow5':6\rightarrow6'$). It is noted that smaller atoms (of type 2) are involved at both ends of the chain sequence, and that both types of atoms are involved in the elastic and plastic accommodation. Such mixed atom type chain-like activity is also seen in the smaller ring-like LSEs as shown in fig.~\ref{FigVis}e, which has an activation energy of $10.55\epsilon$. Finally, fig~\ref{FigVis}f represents an LSE with an activation energy of $8.96\epsilon$ involving 11 atoms. This LSE forms a chain ($1\rightarrow1':2\rightarrow2':3\rightarrow3':4\rightarrow4'$) which almost resembles a straight-line due to its low curvature.

Upon inspection of these figures, the chain like sequence of an LSE generally involves one atom replacing its neighbour (and so on) such that the chain or part of the chain is fully connected (with respect to the red and blue displacement arrows) or one atom can move a previously unoccupied location with another atom doing the same with respect to another atom (and so on) forming a disconnected chain (with respect to the red and blue displacement arrows). The smaller ring like structures of figs.~\ref{FigVis}a and e, fall into the first category and the extended chains fall into both categories. Very low energy LSEs were also visualized (not shown) and these tended to involve just one atom changing its location with minor elastic and plastic accommodation in the surrounding regions. Such LSEs typically have activation energies in the range of less than $\sim5\varepsilon$.

Fig.~\ref{FigVis} demonstrates that LSEs can be spatially extended. To better quantify this observation, a non-dimensional quantity derived from the radius of gyration is used. The radius of gyration (which is also commonly used in molecular applications) in the present context is given by,
\begin{equation}
R_{\mathrm{g}}=\sqrt{\frac{1}{N_{\mathrm{LSE}}}\sum_{k=1}^{N_{\mathrm{LSE}}}\left(\mathbf{r}_{k}-\mathbf{r}_{\mathrm{cop}}\right)^{2}},
\end{equation}
where $N_{LSE}$ is the number of atoms involved in the LSE, $\mathbf{r}_{k}$ is the position of the concerned atom and $\mathbf{r}_{\mathrm{cop}}$ is the centre of position of the LSE. The dimensionless shape number is then defined as $R_{\mathrm{g}}/R_{\mathrm{max}}$, where $R_{\mathrm{max}}$ is the maximum distance from the centre-of-position of an atom within the LSE. The shape number is computed using the initial positions of the atoms involved in the LSEs. A higher displacement cut-off of $0.3 \AA$ was used to exclude the surrounding accommodating atoms (which have relatively smaller displacements), thus ensuring only the central atoms within the LSE are considered. 

\begin{figure}
\centering
\includegraphics[scale=0.75]{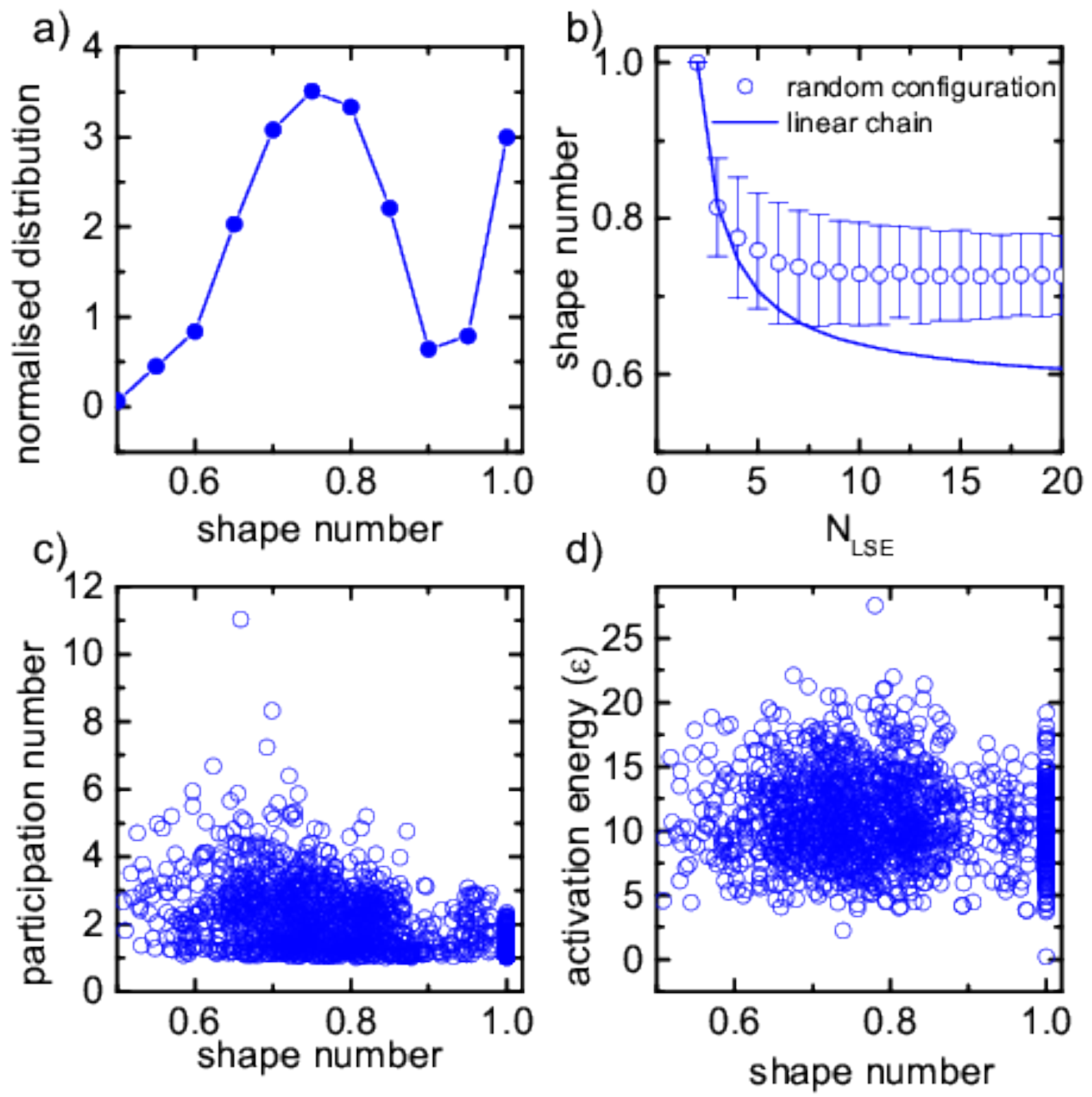}
\caption{\label{FigShapeNumber} a) the shape number distribution derived from all identified LSEs, b) the scatter plot of shape number with participation number. c) shows the relationship between LSE size and shape number for the two limiting geometries of a linearly extended LSE and a randomly spatially distributed LSE. d) the scatter plot of shape number with activation energy.}
\end{figure}

Fig.~\ref{FigShapeNumber}a displays the histogram of shape numbers derived from all of the identified LSEs. To understand this figure, some limiting cases of the shape number formalism are first considered. When only one central atom is involved in an LSE, the shape number becomes indeterminate and when two atoms are involved the shape number becomes its largest value of unity. An LSE consisting of an arbitrary number of atoms all situated on the surface (perimeter) of a sphere (circle) also gives a value of unity. Fig.~\ref{FigShapeNumber}b displays the shape number as a function of $N_{\mathrm{LSE}}$ for the cases of a perfect linear chain and a spherical volume in which the positions of the atoms are chosen randomly. The data corresponding to the latter case also includes the converged variance. In fig.~\ref{FigShapeNumber}a, the peak at a shape number of unity is therefore either due to a strong population of perfect rings containing an arbitrary number of atoms or the many two atom LSEs discussed previously. Inspection of fig.~\ref{FigShapeNumber}c, which plots the scatter diagram between shape number and participation number, demonstrates the latter case, where there exists a dense line of LSEs with a participation number of approximately two at a shape number of one. For larger LSEs fig.~\ref{FigBED}c shows that typical participation numbers are between 3 and 5. In this regime, the shape number for the limiting cases of fig.~\ref{FigShapeNumber}b admit shape numbers ranging between 0.7 and 0.85, which is precisely the location of the central peak in fig.~\ref{FigShapeNumber}a. Fig.~\ref{FigShapeNumber}c shows that LSEs consisting of greater than five atoms have shape numbers spanning the entire range of possible values indicating that both extended and more compact LSE structures are contributing to the peak, a conclusion that is seen in fig.~\ref{FigVis}. The enhanced tail at low shape numbers does suggest larger extended LSE chains. Importantly, in most cases an LSE may be characterized by an approximate sequence of atoms replacing atoms rather than a random rearrangment of atomic positions.  Fig.~\ref{FigShapeNumber}d displays a scatter graph between shape number and activation energy showing that there is little correlation between the spatial extension of an LSE and its barrier energy.

\section{Discussion} \label{SecDiscussion}

The results of sec.~\ref{SecResults} suggest that the location of an LSE is only weakly correlated with the local structural features of those atoms involved. For the LJ system considered, the only non-negligible correlation is that the smaller atoms of type 2, are more often involved than the larger atoms of type 1, particularly when the LSEs consist of only a few atoms and are at the lower range of the activation energy spectrum. Despite the strong scatter, this latter aspect suggests a rather intuitive scenario where, because type two atoms generally involve less negative bond energies, the breaking of bonds that must occur in an LSE requires less energy and therefore the activation energy should be lower. Indeed this appears to be more important than local Voronoi volume since in fig~\ref{FigLAQ}b, the volume LSE weighted distribution exhibits only a central peak structure not located at volumes typical of type 2 atoms, whereas the local cohesive energy LSE distribution clearly correlates with the type 2 unweighted peak (fig~\ref{FigLAQ}a). The remaining, somewhat weaker, correlation with activation energy is that a small or negative lowest local Kelvin eigen-shear tends to have a low activation energy. Again, an intuitive result since a small or negative Kelvin eigen-shear indicates a shallow potential energy minimum and therefore a smaller activation barrier. In other words, the LSEs occurring in softer regions tend to have lower activation energies.

The atomistic visualization shown in sec.~\ref{SecAV} generally demonstrates LSEs to be a sequence of atoms that successively replace each others approximate location, with the surrounding atoms accommodating such movement through either elastic or plastic distortion. This appears to be a general result, although the spatial extension of the atomic sequence can be quite diverse, ranging from an almost linear extension, to strongly curved  and closed ring-like structures (for the smaller LSEs). Although those central atoms within the chain show no obvious decrease in their own local Voronoi volume, it is of interest to investigate whether nearby free volume is correlated with their existence. This is motivated by the original assumption of Spaepen in his thermally activated free-volume theory~\cite{Spaepen1977}. To determine the spatial location of free volume within the computer generated sample, the simulation cell is filled with a fine regular cubic mesh of points, at a spacing much smaller than the typical inter-atomic distance of $\sim\sigma$. Those mesh points that have a distance to the nearest atom greater than $R_{\mathrm{max}}$, and which are connected to each other, will then define the spatial extent of a region of free local volume. $R_{\mathrm{max}}$ cannot be too small since then the normal interstitial regions, which span the entire simulation cell, will be identified. The parameter should also not be too large since then no free volume will be identified. Such a method has been used to identify free volume in grain boundaries~\cite{VanSwygenhoven2008}. Using a value of  $R_{\mathrm{max}}=0.76\sigma$, fig.~\ref{FigCOP} displays the identified regions as green balls. This figure shows that the computer generated sample contains a few regions in which local free volume is above the normal background of interstitial regions.

\begin{figure}
\centering
\includegraphics[scale=0.7]{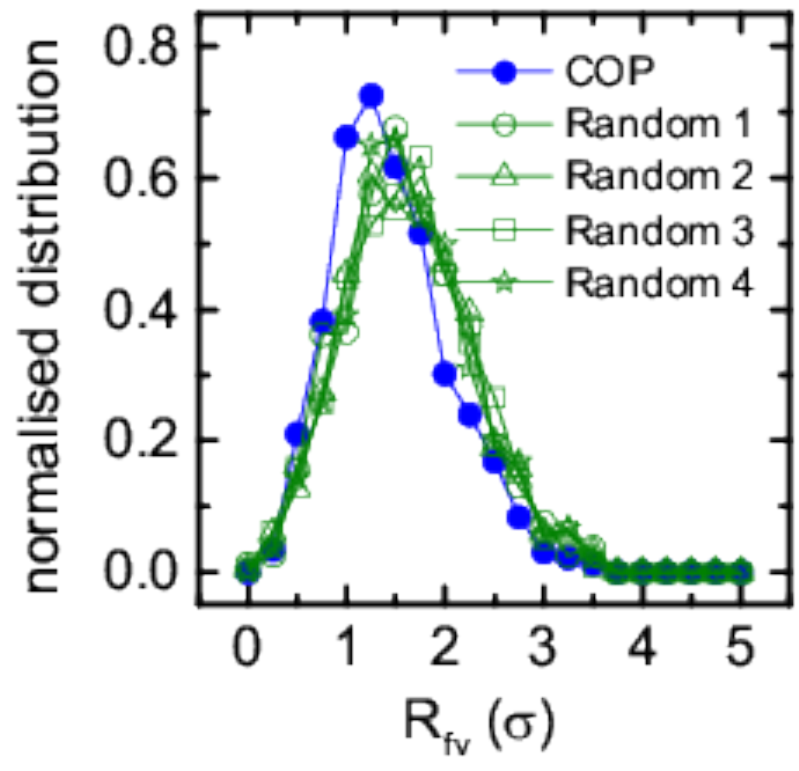}
\caption{\label{FigFreeVolume}Distribution of minimum distances between the centre-of-position of an LSE and free-volume in the sample. For comparison similar distributions are shown for the case when LSEs are distributed uniform randomly throughout the simulation cell.}
\end{figure}

Fig.~\ref{FigFreeVolume} now shows a histogram of, $R_{\mathrm{fv}}$, the nearest LSE centre-of-position (the red coloured balls in fig.~\ref{FigCOP}) to all identified free volume (the green coloured balls in fig.~\ref{FigCOP}). Also shown are histograms of four random realizations of LSE center-of-positions derived from a uniform distribution within the simulation cell. Inspection of this figure shows a slight bias of the ART$n$ identified LSEs to be closer to free volume than that of an entirely randomly located LSEs. Thus there exists some correlation between the location of an LSE and nearby free-volume.

The framework of fast $\beta$ and slower $\alpha$ structural transformations, originally developed for under-cooled liquids is now often applied to the regime of amorphous solids~\cite{Harmon2007,Yu2012,Yu2013,Derlet2011,Derlet2012a,Derlet2013,Derlet2013b}. Indeed, in the work of Harmon {\em et al} \cite{Harmon2007} multiple microscopic $\beta$ structural transformations (which are assumed to be reversible) underly the emergence of irreversible $\alpha$ transformations in the form of a less local release of elastic energy. In more recent work~\cite{Yu2012}, which attempts to explain dynamical-mechanical-spectroscopy data, some of these authors have postulated $\beta$ structural transformations to consist of atomic chains of smaller atoms, rather like those encountered in the present work. On the other hand structural transformations that involve both the small and large atoms, tend to reflect $\alpha$ transformations --- a rather intuitive picture since movement of the larger atoms will tend to involve more atoms due to accommodation issues and therefore be inherently less local. The usage of the terminology of $\alpha$ and $\beta$ transformations in the regime of the amorphous solid is an interesting development given that from the under-cooled liquid perspective the $\alpha$ relaxations are assumed to be frozen out below the glass transition temperature~\cite{Angell2000}. How this freezing occurs, and how far it extends to temperatures and affects plasticity below the glass transition has recently been considered by one of the present authors from the perspective of a thermal activation theory of deformation~\cite{Derlet2012a,Derlet2013,Derlet2013b}.

\begin{figure}
\centering
\includegraphics[scale=0.7]{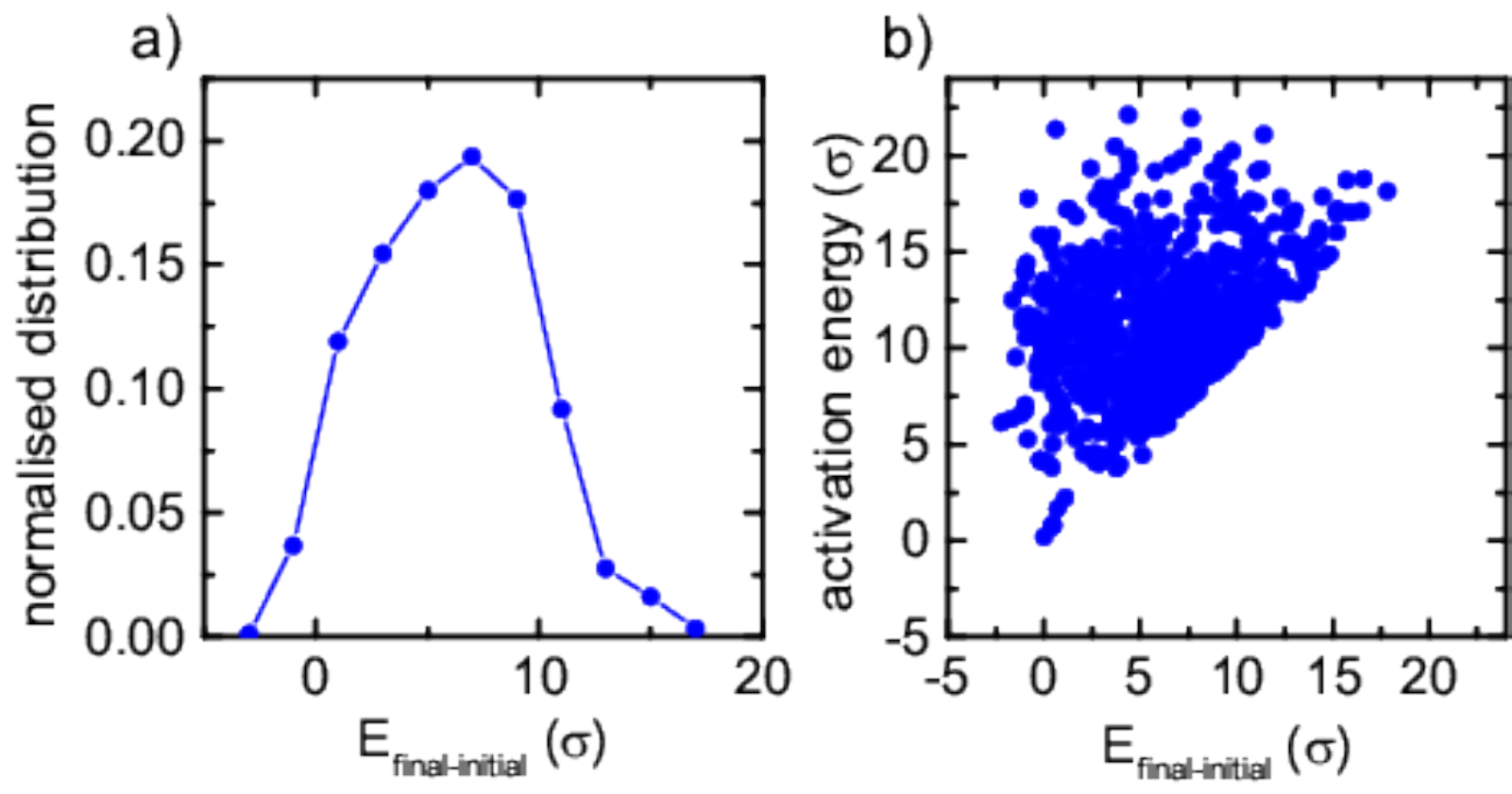}
\caption{\label{FigAorB} a) Normalized distribution of the change in energy between the initial and final state configurations and b) a scatter plot of this energy difference with the corresponding activation energy.}
\end{figure}

Within the above framework a relevant question is, to which class of relaxation processes ($\alpha$ or $\beta$) should the identified LSEs belong. Fig.~\ref{FigAorB}a displays a histogram of the change in energy between the initial and final atomic configurations found by ART$n$. In most cases this energy is positive, with a few LSEs leading to a decrease in energy and therefore a more stable atomic configuration than the initial configuration reached by dynamical atomistic simulations. Given that only LSEs are considered which have a direct path between the initial and activated states (that is, there exists no intermediate stable configuration), fig.~\ref{FigAorB}a suggests that the initial configuration is in the basin of a much larger PEL valley and therefore in the valley of the $\alpha$ landscape. From this context, the ART$n$ method appears to be probing primarily the $\beta$ PEL involving the first LSE stage that would begin the atomic configuration's journey out of its current $\alpha$ mega-basin. Fig.~\ref{FigAorB}b shows the scatter plot between the change in energy between the initial and final atomic configuration and the corresponding activation energy. The plot demonstrates the obvious fact that an activated energy cannot be less than the final state energy for LSEs that are directly connected to the initial state. The figure also reveals that those final states that have an energy less than the initial state are separated by the full spectrum of possible activation energies with only a very few having quite small activation energies. Generally, little correlation is seen apart from the observation that both energy scales are comparable demonstrating that, if the assumed surrounding $\alpha$ energy landscape does exist, the underlying ``ripple'' $\beta$ energy scale is that of the LSE energy scale. Thus it could be said that $\beta$ processes are not so sensitive to their local environment and that no such statement can be made for the $\alpha$ processes. Fig.~\ref{FigVis} shows however that identifying LSEs as $\beta$ processes has the consequence that bonds are broken for the latter --- a result that is different from the view point that only $\alpha$ processes involve the breaking of bonds (see for example ref.~\cite{Rodney2011} and references therein). Clearly further work is needed to confirm this picture. 

Whilst the results found here are not sensitive to a LJ sample realization, as well as the quench rate, the sample size is rather small containing only 1728 atoms --- a cell size side length of only $\simeq8\sigma$. The computational load of the ART$n$ algorithm makes it difficult to apply the technique to much larger systems. Application of the ART$n$ method to similar LJ systems containing 34,000 atoms (corresponding to side lengths of $\simeq30\sigma$) do reveal LSEs of similar size to that encountered here suggesting that the LSEs accessible to ART$n$ are at the microscopic length scale~\cite{Derlet2013a}. Noting that ART$n$ converges to nearby minima, this is entirely compatible with the $\alpha-\beta$ potential energy landscape in which the larger and irreversible $\alpha$ transformation landscape is rippled with those of the underlying $\beta$ transformation landscape. The usage of more realistic inter-atomic multi-component potentials is not expected to fundamentally change the current results, although some fine details of the LSE structure unique to the potential/system are expected to occur which are not present in the considered model LJ studied here. Finally it is emphasized that it is assumed that ART$n$ provides an unbiased probe to nearby saddle-point configurations within the potential energy landscape --- an assumption that underlies all previous work applying ART$n$ to structural glasses~\cite{Rodney2009a,Rodney2009b,Kallel2010,Koziatek2013}. 

\section{Conclusions} 

In the present work, the ART$n$ method has been used to identify Local Structural Excitations (LSEs) in three-dimensional model glass samples that have been characterised in terms of their Local Atomic Quantities (LAQ). Taking advantage of the localised displacement fields of the LSEs, the mean atomic quantity of an LSE is computed via a displacement weighting technique and its distribution compared with total sample and atom type resolved distributions. It has been found that only a weak correlation exists between the local atomic environment and where an LSE occurs, and its activation energy. In particular, it is found that smaller atoms are more often involved and that those with lower activation energy tend to occur in softer potential energy regions. The origin of this appears to lie in the weak bond energy between such atoms. In general, however, LSEs identified via the ART$n$ method occur throughout the sample, and in large number, with a slight bias to be near regions of free volume. Atomistic visualization of individual LSEs reveal them to consist of chain-like structures involving the successive replacement of one atom with that of a nearest neighbour atom and a surrounding accommodation mechanism involving both elastic and plastic distortion.

\end{document}